%% file: nonlinear.tex
\documentclass[twocolumn]{IEEEtran} 
   \usepackage{epsfig}
   \usepackage{color}
   \usepackage{latexsym}
   \usepackage{pslatex}
  \usepackage{algorithm}
   \usepackage{algorithmic}

\newtheorem{lemma}{Lemma}
\newtheorem{theorem}{Theorem}

\begin{document}

\title{Synthesis of Binary $k$-Stage Machines} 
\author{Elena~Dubrova, \IEEEmembership{Member, IEEE}%
\thanks{E. Dubrova is with the Royal Institute of Technology (KTH), Stockholm, Sweden.}}

\maketitle

\begin{abstract}
An algorithm for constructing a shortest binary $k$-stage machine 
generating a given binary sequence is presented. This algorithm 
can be considered as an extension of Berlekamp-Massey algorithm to the non-linear case. 
\end{abstract}


\section{Introduction}

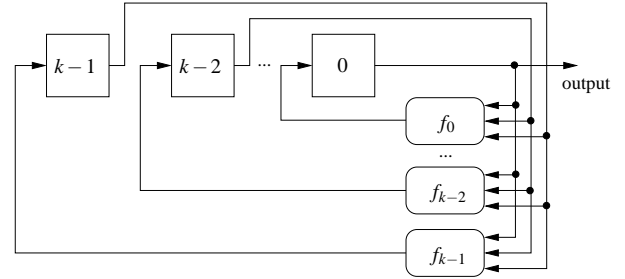
\begin{figure}[t!]
\begin{center}
\resizebox{0.9\columnwidth}{!} {\input{binary_machine.xfig.pstex_t}}
\caption{ A binary $k$- stage machine.}\label{bin_machine}
\end{center}
\end{figure}

In his seminal book~\cite{Golomb_book} Golomb described an extended version of the traditional
feedback shift register, shown in Figure~\ref{bin_machine}. He called such a device {\em binary}
$k$-{\em stage machine}. Each stage $i \in \{0,1,\ldots,k-1\}$ has its own next state function $f_i$.
Both feedback and feedforward connections are allowed.  

In this paper, we address the problem of constructing a binary
$k$-stage machine with the minimum $k$ generating a given
binary sequence. We present a synthesis algorithm and derive the exact lower bound on
$k$. Our work can be considered as an extension of Berlekamp-Massey
algorithm~\cite{Ma69} to the non-linear case. 

For the traditional Non-Linear Feedback Shift Registers (NLFSRs), 
the problem of finding a shortest NLFSR generating a
given binary sequence has been considered in~\cite{Ja91,RiK05,RiKK05}
and~\cite{LiKK07}.  

\section{Preliminaries} 

A {\em binary sequence} $A$ of length $n$ is an $n$-tuple $(a_0, a_1, \ldots, a_{n-1})$
where $a_i \in \{0,1\}$ for $i \in \{0,1,\ldots,n-1\}$. The {\em Hamming weight} of a binary sequence $A$,
denoted by  $wt(A)$,  is the number of 1s in $A$. A binary sequence $A$ of length $n$ is  {\em balanced}
if $wt(A) = n-wt(A)$.

For a Boolean function $f: \{0,1\}^n \rightarrow \{0,1\}$, the {\em support} of $f$ is defined by 
\[
\Omega_{f} = \{x \in \{0,1\}^n: f(x) = 1\}.
\]

The {\em algebraic normal form} (ANF) of a Boolean function $f$ is a polynomial in $GF(2)$ of type
\[
f(x_0, \ldots, x_{n-1}) = \sum_{i=0}^{2^n-1}  r_i \cdot 
x_0^{i_0} \cdot x_1^{i_1} \cdot \ldots \cdot x_{n-1}^{i_{n-1}},
\]
where $r_i \in \{0,1\}$ and $(i_{n-1} \ldots  i_1 i_0)$ is the binary
expansion of $i$ with $i_0$ being the least significant bit. 

The {\em gate complexity}~\cite{massey_talk}  (or {\em circuit-size complexity}) of a Boolean function $f$ is the smallest 
number of gates in any acyclic circuit computing $f$, given that the 
gates are restricted to have at most two inputs.

A {\em state} of a binary $k$-stage machine is a vector of values of its $k$ stages.

\section{Synthesis Algorithm} \label{syn_alg}

The algorithm presented in this section exploits the property of binary
$k$-stage machines that {\em  any} binary $k$-tuple can be the next state of 
a given current state. Note that, in a traditional NLFSR in the Fibonacci configuration~\cite{Golomb_book}, the next state
overlaps with a current state in $k-1$ positions.
The Galois configuration of NLFSRs, introduced in~\cite{Du09j}, is more flexible.
However, since feedforward connections are not allowed in NLFSRs,
the set of possible next states is still limited.

First, we show how to construct a sequence of integers 
whose least significant bits follow a given aperiodic binary sequence of length $n$.

Let $B = (0,2,4,\ldots)$ be an infinite vector of all even non-negative integers starting from 0.  
Let $C = (1,3,5,\ldots)$ be an infinite vector of all odd positive integers starting from 1.  
We denote by $b_i$ and $c_i$ be the $i$th elements of $B$ and $C$, respectively,
for $i \in \{0,1,2\ldots\}$.

Let $N_0 = 0$ and $N_1 = 0$.  Given an aperiodic binary sequence $A$ of length $n$, for every $i$ from 0 to $n-1$, we repeat the following: 

\begin{itemize}
\item[]
If $a_i = 0$, then assign $s_i = b_{N_0}$ 
and increment $N_0$ by one. Otherwise, assign $s_i =  c_{N_1}$ 
and increment $N_1$ by one. 
\end{itemize}

The algorithm described above is summarized as Algorithm~\ref{alg1}.
Its worst-case time complexity is $O(n)$.

Let $S = (s_0, s_1, \ldots, s_{n-1})$ be a sequence constructed by the Algorithm~\ref{alg1}.
Each integer $s_i \in S$ can be represented as a binary expansion
$(s_{i_{k-1}}, s_{i_{k-2}}, \ldots, s_{i_0}) \in \{0,1\}^k$ where 
$k$ is the number of bits needed to represent the largest integer of $S$
and $s_{i_0}$ is the least significant bit of the expansion.
We interpret each $k$-tuple $(s_{i_{k-1}}, s_{i_{k-2}}, \ldots, s_{i_0})$ as a state
of a binary $k$-stage machine.
By construction, $s_{i_0} = a_i$ for all $i \in \{0,1,\ldots,n-1\}$.

Next, we define a mapping $s_i \mapsto s_{i+1}$, for all $i \in
\{0,1,\ldots,n-1\}$, where $''+''$ is modulo $n$. This mapping assigns 
$s_{i+1}$ to be the next state of a current state $s_i$ of a binary $k$-stage machine. Each of $2^k - n$ remaining
states of the binary $k$-stage machine are mapped into the all-0 state. This
implies that they do not contribute any 1s to the supports  of
the next state functions.

The supports of the next state functions 
implementing the resulting mapping are derived as follows.
Initially $\Omega_{f_j} = \emptyset$, for all $j \in \{0,1,\ldots,k-1\}$. For every $i$ from 0 to $n-1$,
we repeat the following:

\begin{itemize}
\item[]
For every $j$ from 0 to $k-1$: If $s_{(i+1)_j} = 1$, where $''+''$ is modulo $n$, then 
\[
\Omega_{f_j} = \Omega_{f_j} \cup \{(s_{i_{k-1}},s_{i_{k-2}},\ldots, s_{i_0})\}.
\]
\end{itemize}

The algorithm described above is summarized as Algorithm~\ref{alg2}.
Its worst-case time complexity is $O(n \cdot k)$.

\begin{algorithm}[t]
\caption{Construct a sequence of non-negative integers whose least significant bits
follow an aperiodic binary sequence $A = (a_0, a_1, \ldots, a_{n-1})$.}
\label{alg1}
\begin{algorithmic}[1]
\STATE $B = (0,2,4,\ldots)$; /*even non-negative integers*/
\STATE $C = (1,3,5,\ldots)$; /*odd positive integers*/
\STATE $N_0 := 0$;
\STATE $N_1 := 0$;
\FOR{every $i$  from 0 to $n-1$}
\IF{$a_i = 0$}
\STATE $s_i := b_{N_0}$; /*$b_i$ is the $i$th element of $B$ */
\STATE $N_0 := N_0 + 1$;
\ELSE
\STATE $s_i := c_{N_1}$; /*$c_i$ is the $i$th element of $C$ */
\STATE $N_1 := N_1 + 1$;
\ENDIF
\ENDFOR
\STATE Return $S := (s_0, s_1, \ldots, s_{n-1})$;
\end{algorithmic}
\end{algorithm}

\begin{theorem} \label{th1}
The algorithm presented in this section constructs a binary $k$-stage machine 
generating a finite aperiodic binary sequence $A$ where $k$ is given by
\begin{equation} \label{bound}
k = max(\lceil log_2 wt(A) \rceil, \lceil log_2 (n-wt(A)) \rceil) + 1,
\end{equation}
where $n$ is the length of $A$.
\end{theorem}
{\bf Proof:}  
When the Algorithm~\ref{alg1} terminates, $N_1 = wt(A)$.
Since $A$ is aperiodic, we have $0 < wt(A) < n$.
Therefore,  the largest odd integer used from $C$ is $2 wt(A) - 1$.
The binary expansion of this odd integer has $\lceil log_2 wt(A) \rceil+ 1$ bits.
Similarly, when the Algorithm~\ref{alg1} terminates, we have $N_0 = n-wt(A)$.
The largest even integer used from $B$ is $2 (n-wt(A)) - 2$.
The binary expansion of this even integer has $\lceil log_2 (n-wt(A)) \rceil + 1$ bits.
\begin{flushright}
$\Box$
\end{flushright}

The following property trivially follows from the Theorem~\ref{th1}.

\begin{lemma} 
If $A$ is balanced, then (\ref{bound}) reduces to
\[
k=  \lceil log_2 n \rceil.
\]
\end{lemma} 

\begin{algorithm}[t!]
\caption{Construct the next state functions for a binary $k$-stage machine
which follows the sequence of states $S = (s_0, s_1, \ldots, s_{n-1})$, $s_i \in \{0,1\}^k$.}
\label{alg2}
\begin{algorithmic}[1]
\FOR{every $j$  from 0 to $k-1$}
\STATE $\Omega_{f_j} = \emptyset$;
\ENDFOR
\FOR{every $i$  from 0 to $n-1$}
\FOR{every $j$  from 0 to $k-1$}
\STATE /*Each $s_i \in S$ is of type $(s_{i_{k-1}}, s_{i_{k-2}}, \ldots, s_{i_0})  \in \{0,1\}^k$*/
\IF{$s_{(i+1)_j} = 1$} 
\STATE $\Omega_{f_j} = \Omega_{f_j} \cup \{(s_{i_{k-1}},s_{i_{k-2}},\ldots, s_{i_0})\}$;
\ENDIF
\ENDFOR
\ENDFOR
\STATE Return $(f_0, f_1, \ldots, f_{k-1})$;
\end{algorithmic}
\end{algorithm}

As an example, consider the following sequence of length $n = 19$ taken from the Example V.1 in~\cite{LiKK07}:
\[
A = (0011011100101110110).
\]
It was shown in~\cite{LiKK07} that the shortest NLFSR generating this sequence has 7
stages. Below we show that the same sequence can be
generated using a binary machine with 5 stages. This comes as no surprise, since a binary machine is more general than an NLFSR. Using the Algorithm~\ref{alg1}, we construct the following sequence 
of integers whose least significant bits
follow $A$:
\[
S = (0,2,1,3,4,5,7,9,6,8,11,10,13,15,17,12,19,21,14).
\]
By applying the Algorithm~\ref{alg2} to $S$, we get the following supports for the next state functions:
\[
\begin{array}{l}
\Omega_{f_4} =  \{(01100),(01111),(10011)\}\\[1mm]
\Omega_{f_3}  =   \{(00110),(00111),(01000),(01010),(01011),\\
~~~~~~~~  (01101),(10001),(10101)\}\\[1mm]
\Omega_{f_2}  =   \{(00011),(00100),(00101),(01001),(01010),\\
~~~~~~~~  (01101),(10001),(10011),(10101)\}\\[1mm]
\Omega_{f_1}  =   \{(00000),(00001),(00101),01000),01001),\\
~~~~~~~~  (01011),(01100),(01101),(10101)\}\\[1mm]
\Omega_{f_0}  =   \{(00001),(00010),(00100),(00101),(00111),\\
~~~~~~~~ (01000),(01010),(01100),(01101),(01111),\\
~~~~~~~~ (10011)\}\\
\end{array}
\]
These supports have the following ANF expressions:
\[
\begin{array}{l}
f_4 = x_0 x_1 x_3 \oplus x_1 x_2 x_3 \oplus x_1 x_4 \oplus x_0 x_1 x_4 \oplus x_1 x_2 x_4 \oplus x_0 x_1 x_2 x_4\\ ~~~~~ \oplus x_1 x_3 x_4 \oplus x_0 x_1 x_2 x_3 x_4 \\

f_3 = x_0 x_2 \oplus x_1 x_2 \oplus x_0 x_1 x_2 \oplus x_0 x_3 \oplus x_1 x_3 \oplus x_2 x_3 \oplus x_0 x_2 x_3\\ ~~~~~ \oplus x_1 x_2 x_3 \oplus x_4 \oplus x_0 x_4 \oplus x_1 x_4 \oplus x_0 x_1 x_4 \oplus x_0 x_2 x_4 \oplus x_1 x_2 x_4\\
~~~~~  \oplus x_0 x_1 x_2 x_4 \oplus x_3 x_4 \oplus x_0 x_1 x_3 x_4 \oplus x_2 x_3 x_4 \oplus x_0 x_2 x_3 x_4\\ ~~~~~ \oplus x_1 x_2 x_3 x_4 \\

f_2 = x_1 \oplus x_2 \oplus x_0 x_2 \oplus x_0 x_1 x_2 \oplus x_3 \oplus x_2 x_3 \oplus x_4 \oplus x_0 x_4 \oplus x_1 x_4\\
~~~~~  \oplus x_2 x_4 \oplus x_0 x_2 x_4 \oplus x_1 x_2 x_4 \oplus x_0 x_3 x_4 \oplus x_2 x_3 x_4 \oplus x_1 x_2 x_3 x_4\\
~~~~~  \oplus x_0 x_1 x_2 x_3 x_4 \\

f_1 = 1 \oplus x_1 \oplus x_2 \oplus x_0 x_2 \oplus x_1 x_2 \oplus x_0 x_1 x_2 \oplus x_0 x_1 x_3 \oplus x_2 x_3\\
~~~~~  \oplus x_0 x_2 x_3 \oplus x_1 x_2 x_3 \oplus x_4 \oplus x_1 x_4 \oplus x_2 x_4 \oplus x_1 x_2 x_4 \oplus x_0 x_1 x_3 x_4\\
~~~~~  \oplus x_2 x_3 x_4 \oplus x_1 x_2 x_3 x_4 \oplus x_0 x_1 x_2 x_3 x_4 \\

f_0 = x_0 \oplus x_1 \oplus x_2 \oplus x_0 x_2 \oplus x_0 x_1 x_2 \oplus x_3 \oplus x_1 x_3 \oplus x_2 x_3 \oplus x_1 x_2 x_3\\
~~~~~  \oplus x_0 x_4 \oplus x_1 x_4 \oplus x_0 x_1 x_4 \oplus x_2 x_4 \oplus x_0 x_2 x_4 \oplus x_3 x_4 \oplus x_1 x_3 x_4\\
~~~~~  \oplus x_0 x_1 x_3 x_4 \oplus x_2 x_3 x_4 \oplus x_1 x_2 x_3 x_4 \oplus x_0 x_1 x_2 x_3 x_4 
\end{array}
\]

As we can see, the resulting next state functions have a substantial gate complexity. We can potentially reduce the gate complexity as follows:
\begin{enumerate}
\item By using a different sequence of states to generate $A$. In general, any permutation of even integers from the set $\{0,2,4,\ldots,2(n-wt(A))-2\}$ and any permutation of odd integers from the set $\{1,3,5,\ldots,2wt(A)-1\}$ can be used in the Algorithm~\ref{alg1} instead of vectors $B$ and $C$, respectively, to construct a sequence of integers whose least significant bits follow $A$.
\item By mapping the remaining $2^k-n$ states of the binary $k$-stage machine in a different way.
For example, rather than being mapped into the all-0 state, these states can form another cycle of states.
The resulting binary $k$-stage machine will be branchless.
\end{enumerate}

In general, the problem of constructing a binary $k$-stage machine with the minimum gate complexity 
of next state functions is very hard. It is unlikely that there exists an exact algorithm for solving this problem 
which is feasible for large $n$.

\section{Bound on the Size}

The theorem below shows that the bound given by~(\ref{bound}) is exact.

\begin{theorem} \label{th2}
Given a finite aperiodic binary sequence $A$ of length $n$, any 
binary machine which can generate $A$  has at least $k$ stages, where
$k$ is given by~(\ref{bound}).
\end{theorem}
{\bf Proof:} 
The existence of a binary machine with $k$ stages which can generate $A$
follows from the Theorem~\ref{th1}.  
It remains to prove that no binary $k'$-stage machine with $k' < k$ can generate $A$.

Assume that $k$ is given by (\ref{bound}) and that there exists a binary machine with  
$k^{'}$ stages, $k^{'} < k$, which can generate the same sequence $A$. 

Let $wt(A) \geq n/2$. One one hand, from~(\ref{bound}), we have $k = \lceil log_2 wt(A) \rceil + 1$.
On the other hand, to be able to generate an aperiodic binary sequence $A$, a binary $k'$-stage machine 
must have at least $wt(A)$ distinct states with the least significant bit 1. Therefore, it must have at least 
$k' \geq \lceil log_2 wt(A) \rceil + 1$ stages. This contradict the assumption $k^{'} < k$.

In a similar way, we can come to a contradiction for the case  $wt(A) < n/2$.
Therefore, no binary machine with less than $k$ stages can generate $A$.
\begin{flushright}
$\Box$
\end{flushright}


\section{Conclusion}

We presented an algorithm for constructing a shortest binary
$k$-stage machine generating a given binary sequence. 
Since binary $k$-stage machines are probably the most general extension of NLFSRs,
the lower bound given by the Theorem~\ref{th2} might be useful for estimating non-linear 
complexity of sequences. 

Future work includes finding a heuristic approach for choosing a sequence of states which 
minimizes the gate complexity of the next state functions.


\bibliographystyle{ieeetr}
\bibliography{bib}


\begin{biography} []{Elena Dubrova}
received the Diploma Engineer degree in Computer Science from the Technical University of Sofia, Bulgaria, in 1993, and  the Ph.D. degree in Computer Science from University of Victoria, B.C., Canada, in 1997. Currently she is a professor in Electronic System Design at the School of Information and Communication Technology at Royal Institute of Technology, Stockholm, Sweden. 

She held visiting appointments at the University of New South Wales, Sydney, in 2002, the University of California at Berkeley in 2003, and the University of Queensland in 2005. She has authored over 100 publications in the area of electronic system design. Major contributions include new algorithmic techniques for Boolean decomposition, FPGA technology mapping, and probabilistic verification. Her work has been awarded prestigious prices such as IBM faculty partnership award for outstanding contributions to IBM research and development. Her current research interests include logic synthesis, fault-tolerant computing, formal verification, cryptography, and systems biology. 
\end{biography}

\end{document}

%% file: binary_machine.xfig.pstex_t
\begin{picture}(0,0)%
\includegraphics{binary_machine.xfig.pstex}%
\end{picture}%
\setlength{\unitlength}{3947sp}%
\begingroup\makeatletter\ifx\SetFigFont\undefined%
\gdef\SetFigFont#1#2#3#4#5{%
  \reset@font\fontsize{#1}{#2pt}%
  \fontfamily{#3}\fontseries{#4}\fontshape{#5}%
  \selectfont}%
\fi\endgroup%
\begin{picture}(5754,2649)(2389,-5548)
\put(5493,-3577){\makebox(0,0)[lb]{\smash{{\SetFigFont{12}{14.4}{\rmdefault}{\mddefault}{\updefault}{\color[rgb]{0,0,0}$0$}%
}}}}
\put(2776,-3586){\makebox(0,0)[lb]{\smash{{\SetFigFont{12}{14.4}{\rmdefault}{\mddefault}{\updefault}{\color[rgb]{0,0,0}$k-1$}%
}}}}
\put(3976,-3586){\makebox(0,0)[lb]{\smash{{\SetFigFont{12}{14.4}{\rmdefault}{\mddefault}{\updefault}{\color[rgb]{0,0,0}$k-2$}%
}}}}
\put(6451,-4120){\makebox(0,0)[lb]{\smash{{\SetFigFont{12}{14.4}{\rmdefault}{\mddefault}{\updefault}{\color[rgb]{0,0,0}$f_0$}%
}}}}
\put(6376,-4786){\makebox(0,0)[lb]{\smash{{\SetFigFont{12}{14.4}{\rmdefault}{\mddefault}{\updefault}{\color[rgb]{0,0,0}$f_{k-2}$}%
}}}}
\put(6376,-5386){\makebox(0,0)[lb]{\smash{{\SetFigFont{12}{14.4}{\rmdefault}{\mddefault}{\updefault}{\color[rgb]{0,0,0}$f_{k-1}$}%
}}}}
\end{picture}%

%% file: nonlinear.bbl
\begin{thebibliography}{1}

\bibitem{Golomb_book}
S.~Golomb, {\em Shift Register Sequences}.
\newblock Aegean Park Press, 1982.

\bibitem{Ma69}
J.~Massey, ``Shift-register synthesis and {BCH} decoding,'' {\em IEEE
  Transactions on Information Theory}, vol.~15, pp.~122--127, 1969.

\bibitem{Ja91}
C.~J.~A. Jansen, ``The maximum order complexity of sequence ensembles,'' {\em
  Lecture Notes in Computer Science}, vol.~547, pp.~153--159, 1991.
\newblock Adv. Cryptology-Eupocrypt'1991, Berlin, Germany.

\bibitem{RiK05}
P.~Rizomiliotis and N.~Kalouptsidis, ``Results on the nonlinear span of binary
  sequences,'' {\em IEEE Transactions on Information Theory}, vol.~51, no.~4,
  pp.~1555--5634, 2005.

\bibitem{RiKK05}
P.~Rizomiliotis, N.~Kolokotronis, and N.~Kalouptsidis, ``On the quadratic span
  of binary sequences,'' {\em IEEE Transactions on Information Theory},
  vol.~51, no.~5, pp.~1840--1848, 2005.

\bibitem{LiKK07}
K.~Limniotis, N.~Kolokotronis, and N.~Kalouptsidis, ``On the nonlinear
  complexity and {L}empel-{Z}iv complexity of finite length sequences,'' {\em
  IEEE Transactions on Information Theory}, vol.~53, no.~11, pp.~4293--4302,
  2007.

\bibitem{massey_talk}
J.~Massey, ``The difficulty with difficulty.''
\newblock EUROCRYPT '96 IACR Distinguished Lecture.

\bibitem{Du09j}
E.~Dubrova, ``A transformation from the {F}ibonacci to the {G}alois {NLFSR}s,''
  {\em IEEE Transactions on Information Theory}, vol.~55, pp.~5263--5271,
  November 2009.

\end{thebibliography}
